\documentstyle[12pt]{article}
\def\ba{\begin{eqnarray}}
\def\ea{\end{eqnarray}}

\def\lb{\label}
\def\be{\begin{equation}}
\def\ee{\end{equation}}

\begin{document}

$\ $
\vskip 2cm

\begin{center}
{\Large \bf On
Baxterized Solutions of Reflection Equation and
Integrable Chain Models}
\end{center}


\begin{center}
\large{A.P.\,Isaev${}^{a}$ and O.V.\,Ogievetsky${}^{b}$ }
\end{center}

\begin{center}
${}^{a}$ Bogoliubov Laboratory of Theoretical Physics,
Joint Institute for Nuclear Research, \\
Dubna, Moscow region 141980, Russia \\
E-mail: isaevap@theor.jinr.ru
\\
\vspace{0.3cm}
${}^{b}$ Center of Theoretical Physics\footnote{Unit\'e Mixte de Recherche
(UMR 6207) du CNRS et des Universit\'es Aix--Marseille I,
Aix--Marseille II et du Sud Toulon -- Var; laboratoire affili\'e \`a la FRUMAM (FR 2291)}, Luminy,
13288 Marseille, France \\
and P. N. Lebedev Physical Institute, Theoretical Department,
Leninsky pr. 53, 117924 Moscow, Russia \\
E-mail: oleg@cpt.univ-mrs.fr
\end{center}

\vspace{2cm}

\noindent
{\bf Abstract.} Non-polynomial baxterized solutions of
reflection equations associated with
affine Hecke and affine Birman-Murakami-Wenzl algebras
are found. Relations to integrable
spin chain models with nontrivial boundary conditions are discussed.

\vskip 1cm
\section*{Introduction}
\setcounter{equation}0

A reflection equation was introduced by Cherednik in \cite{Cher}
as an additional factorization condition for
a boundary S-matrix
which describes the motion of relativistic particles on a half line.
In the context of 2-dimensional integrable
field theories with boundaries the reflection equations
and their solutions were investigated in \cite{GZam}.
These equations and their solutions (boundary $K$-matrices)
were also used in \cite{Skl} for a formulation of integrable spin chains with
nonperiodic boundary conditions and studied in many
subsequent papers (see, e.g., \cite{MN}, \cite{Arn}, \cite{DoiLevi} and references therein).

\medskip
To investigate reflection equations
and their solutions, it is convenient to use their universal formulation in terms
of generators of a group algebra of a braid group and its quotients
like Hecke $H_{M}$ or Birman-Murakami-Wenzl $BMW_M$ algebras. In this approach, solutions
of the Yang-Baxter equation are represented as elements of the above algebras
(baxterized elements), while solutions of the reflection equations are
expressed in terms of generators of affine extensions of $H_{M}$ or $BMW_M$.
Many papers are devoted to investigations of baxterized solutions of the reflection equation in
the Hecke algebra case (see, e.g., \cite{MudK}, \cite{DoiLevi} and references therein).
Conversely, not so much is known about baxterized solutions of the reflection equation in
the Birman-Murakami-Wenzl algebra case (see, however, \cite{RO}).

\medskip
In this paper, new baxterized solutions of the reflection equation of the Hecke
and Birman-Murakami-Wenzl types are found. These solutions are rational functions
of affine generators
which automatically satisfy a unitarity condition. For the cyclotomic quotients of the affine
Hecke and BMW algebras, when affine generators
satisfy a polynomial equation of a finite degree, these solutions can be reduced to
polynomial ones. In particular, for the cyclotomic affine Hecke algebra,
we reproduce solutions obtained recently in \cite{MudK}.

\smallskip
In the last Section we discuss applications of the reflection
equation solutions to the formulation of the integrable chain systems with
nontrivial boundary conditions.

\section{Solutions of reflection equations \\ for the Hecke algebra}\lb{heckesection}
\setcounter{equation}0

A braid group ${\cal B}_{M+1}$ is generated by elements $\sigma_i$ $(i=1, \dots M)$
subject to relations:
\be
\label{braidg}
\sigma_i \, \sigma_{i+1} \, \sigma_i =
\sigma_{i+1} \, \sigma_i \,  \sigma_{i+1} \; , \;\;\;
\sigma_{i} \,  \sigma_{j} = \sigma_{j} \,  \sigma_{i} \;\;
{\rm for} \;\; |i-j| > 1 \; .
\ee

An $A$-Type Hecke algebra $H_{M+1}\equiv H_{M+1}(q)$ (see, e.g., \cite{Jon1} and references therein) is
a quotient of the group algebra of the braid group ${\cal B}_{M+1}$
by a Hecke relation
\be
\label{ahecke}
\sigma^2_i - 1 = \lambda \,  \sigma_i \; , \;\; (i = 1, \dots , M) \; ,
\ee
where $\lambda := (q -q^{-1})$ and $q \in {\bf C} \backslash \{0\}$ is a parameter.

\vskip .2cm
Let $x \in {\bf C}$ be a {\it spectral} parameter.
The baxterized elements \cite{Jimb1}
\be
\lb{baxtH}
\sigma_n(x) : =  \sigma_n - x \sigma_n^{-1} \in H_{M+1} \; ,
\ee
solve the Yang-Baxter equation:
\be
\lb{ybeH}
\sigma_n(x) \, \sigma_{n-1}(xy) \, \sigma_n(y) =
\sigma_{n-1}(y) \, \sigma_n(xy) \, \sigma_{n-1}(x)  \; .
\ee
Let $a$ be any solution of the equation $ a-\frac{1}{a} = \lambda$; that is, $a$ equals $q$ or $(-q^{-1})$.
Then, for $x\neq aq,-aq^{-1}$,
the elements (\ref{baxtH}) can be represented in the form
\be
\lb{baxtH1}
\sigma_n(x) = \left(a \, x^{-1} - a^{-1}\right) \frac{\sigma_n+a \, x}{\sigma_n + a\, x^{-1}}  \; .
\ee
The element (\ref{baxtH1}) does not depend on the choice of $a=\pm q^{\pm 1}$.
The normalized element
\be
\lb{baxtH2}
\widetilde{\sigma}_n (x;a) := \frac{1}{a \, x - a^{-1}} \, \sigma_n(x) \; ,
\ee
satisfies the unitarity condition
$\widetilde{\sigma}_n (x;a) \, \widetilde{\sigma}_n(x^{-1};a) =1$.

\medskip
An affine Hecke algebra $\hat{H}_{M+1}$ (see, e.g., \cite{ChPr}, Chapter 12.3)
is an extension of the Hecke algebra $H_{M+1}$. The algebra  $\hat{H}_{M+1}$ is generated
by elements
$\sigma_i$ $(i=1, \dots , M)$
of $H_{M+1}$ and affine generators $y_k$ $(k=1, \dots , M+1)$
which satisfy:
\be
\lb{afheck}
y_{k+1} = \sigma_k \, y_k \, \sigma_k \; , \;\;\; y_k \, y_j = y_j \, y_k \; , \;\;\;
 y_j \, \sigma_i  =  \sigma_i \, y_j \;\; (j \neq i,i+1) \; .
\ee
The elements $\{ y_k \}$ generate a commutative subalgebra in $\hat{H}_{M+1}$ while
symmetric functions of $y_k$ form a center in
$\hat{H}_{M+1}$.

\vskip .2cm
The aim of this section is to find solutions $y_n(x)$ of the {\it reflection equation}:
\be
\lb{reflH}
\sigma_n \left(x \, z^{-1}\right) \, y_n(x) \, \sigma_n(x \, z) \, y_n(z) =
y_n(z) \, \sigma_n(x \, z) \, y_n(x) \,  \sigma_n \left(x \, z^{-1}\right) \; ,
\ee
where $y_n(x)$ is an element of the affine Hecke algebra and is
a function of the spectral parameter $x$. We call $y_n(x)$ a
{\it local} solution of (\ref{reflH}) if $[y_n(x), \, \sigma_i]=0$ $(\forall\ i \neq n-1,n)$.
An additional requirement on the solution
$y_n(x)$ is the unitarity condition: $y_n(x) y_n(x^{-1}) = 1$.

\medskip
Consider a subalgebra $\hat{H}^{(n)}_2 \subset \hat{H}_{M+1}$ which is
generated by two elements $\sigma_n,y_n$ (in fact we consider an extension
of $\hat{H}_2^{(n)}$ by formal series in $y_n$).
Symmetric functions of two variables
$y_n$ and $y_{n+1} := \sigma_n y_n \sigma_n$ are central in the algebra $\hat{H}_2^{(n)}$.

\medskip
We make a following Ansatz
\be
\lb{bax2'}
y_n(x) = Z_1(x) + Z_2(x) \,  y_n  \; ,
\ee
where $Z_i(x)$ are central elements in $\hat{H}_2^{(n)}$.

\smallskip
A direct calculation shows that eq.(\ref{reflH}) is satisfied iff
\be
\lb{bax3'}
\frac{Z_1(x)}{Z_2(x)} = \frac{\gamma - z^{(2)} \, x}{x-x^{-1}} \; ,
\ee
where $\gamma$ is a central element in $\hat{H}_2^{(n)}$
independent of the spectral parameter $x$ and
$z^{(2)}=y_n+y_{n+1}$.

Thus a general solution of the reflection equation (\ref{reflH}) within the Ansatz (\ref{bax2'}) is:
\be
\lb{bax4'}
y_n(x) =  Z_2(x) \, \left( y_n + \frac{\gamma - z^{(2)} \, x}{x-x^{-1}}  \right)  \; ,
\ee
where $Z_2(x)$ is an arbitrary central element in $\hat{H}_2^{(n)}$.
In this form the solution is not quite appropriate since it is not obviously local
(e.g., it explicitly depends on $y_{n+1}$). However, one can achieve the locality
(and unitarity) of the solution $y_n(x)$ by specifying the central
element $Z_2(x)$ in (\ref{bax4'}). We now formulate the main result of this section.

\vspace{0.2cm}
\noindent
{\bf Proposition 1.}
{\it In the case of the affine Hecke algebra,
a local unitary solution of the reflection equation (\ref{reflH}) is
\be
\lb{soluH}
y_n(x) = \frac{y_n - \xi \, x }{y_{n} - \xi \, x^{-1}} \; ,
\ee
where $\xi \in {\bf C}$ is an arbitrary parameter.}

\vspace{0.2cm}
\noindent
{\bf Proof.}
The unitarity property: $y_n(x) y_n(x^{-1}) = 1$ of the element
(\ref{soluH}) is obvious. One can rewrite (\ref{soluH}) in the form
\be
\lb{soluH1}
y_n(x) \equiv
Z_2(x) \left( y_n + \frac{\gamma' - x \, z^{(2)} }{x - x^{-1}}  \right)
\ee
where $\gamma'=\xi + (y_n \, y_{n+1})\xi^{-1}$ and
$$
Z_2(x) : = \frac{\xi (x -x^{-1})}{(y_{n} - \xi \, x^{-1}) (y_{n+1} - \xi \, x^{-1})}
\; ,
$$
are central elements in $\hat{H}_2^{(n)}$. The function (\ref{soluH1}) has
the form (\ref{bax4'}) which means that (\ref{soluH})
is the solution of eq.(\ref{reflH}). \hfill $\bullet$

\vspace{0.2cm}
\noindent
{\bf Remark 1.} The solution (\ref{soluH}) is {\it regular}: $y_n(\pm 1)=1$ (see \cite{Skl}).
A general (within the Ansatz (\ref{bax2'})) solution is obtained by a multiplication of
(\ref{soluH}) by an arbitrary scalar function $f(x)$ such that $f(x)f(x^{-1})=1$.

\vspace{0.2cm}
\noindent
{\bf Remark 2.}
If the parameter $\xi$ in (\ref{soluH}) is not equal to 0, $\xi\neq 0$, it can be set to 1 by
an automorphism $\sigma_k \to  \sigma_k$, $y_n \to \xi \, y_n$
of the affine Hecke algebra $\hat{H}_{M+1}$.
In the case $\xi = 0$, the solution (\ref{soluH}) becomes
trivial, $y_n(x)=1$.

\smallskip
The element (\ref{soluH}) stays a solution of (\ref{reflH})
if we substitute instead of $\xi$ any element of $\hat{H}_{M+1}$
central in $\hat{H}_{2}^{(n)}$.

\vspace{0.2cm}
\noindent
{\bf Remark 3.} In a cyclotomic affine Hecke algebra,
the generator $y_1$ satisfies an additional characteristic equation
\be
\lb{chart}
y_1^{m+1} + \sum_{k=0}^{m} \alpha_k y_1^k =0 \; ,
\ee
where $\alpha_k$ are constants and $m$ is a positive integer. The automorphism
mentioned in Remark 2 is obviously broken. In this case
the solution (\ref{soluH}) is equivalent to a polynomial one.
Indeed, the characteristic equation (\ref{chart}) can be
written in the form
\be
\lb{kulm1}
\frac{1}{y_1 - \xi \, x^{-1}} =  \sum_{k=0}^{m} \, b_k(x) \,  y_1^k \; ,
\ee
where ($\alpha_{m+1} := 1$)
$$
b_{m-k}(x) = b_m (x)\, \sum_{r=0}^k \, \alpha_{m-r+1} \, (\xi/x)^{k-r} \, , \;\;
b_m (x)= - \left( \sum_{r=0}^{m+1} \, \alpha_{r} \, \left( \xi/x \right)^{r}  \right)^{-1} \; .
$$
A substitution of (\ref{kulm1}) in (\ref{soluH}) for $n=1$ gives
a polynomial in $y_1$ solution
$$
y_1(x) = (y_1 - \xi \, x ) \, \sum_{k=0}^{m} \, b_k(x) \, y_1^k
$$
\be
\lb{kulm2}
=  b_{m}(x) \, \sum_{k=0}^{m} \, \left( \frac{b_{k-1}(x)}{b_{m}(x)} -
\xi \, x  \, \frac{b_{k}(x)}{b_{m}(x)}- \alpha_k  \right) y_1^k \; , \;\;\;
(b_{-1}(x):=0) \; ,
\ee
of the reflection equation (\ref{reflH}).
The solution (\ref{kulm2}) was obtained in \cite{MudK}.
We stress that the polynomial solution (\ref{kulm2}), which easily follows from the general simple
formula (\ref{soluH}), is valid only in the cyclotomic case and for $n=1$.

\smallskip
For $\xi \neq 0$ the solution (\ref{kulm2}) is called {\it principal} in \cite{MudK}.
Note that in (\ref{kulm2})
one can perform a special limit: $\xi \to 0$, $\alpha_0 \to 0$, $\alpha_0/\xi \to \zeta$,
where $\zeta$ is a constant (in this case, due to eq.(\ref{chart}),
the element $y_1$ is not invertible). As a result we obtain, in this particular case,
following solutions of (\ref{reflH})
$$
y_1(x) = 1 + \frac{x-x^{-1}}{\alpha_1 x^{-1} + \zeta} \, \tilde{y}_1  \; \; \;\;\;
\left( \tilde{y}_1 := y_1^m + \sum_{k=1}^m \alpha_k y_1^{k-1} \; , \;\;\; y_1 \tilde{y}_1 =0 \right) \; ;
$$
these solutions are called {\it small} in \cite{MudK}.

\medskip
Below we explicitly consider few special examples for the polynomial
solutions (\ref{kulm2}) in the cases $m=1,2,3$ and $\xi \neq 0$.

\vspace{0.2cm}
\noindent
\underline{Case $m=1$.}
In this case the characteristic identity (\ref{chart}) for
the affine element $y_1$ is of order 2: $y_1^2 +\alpha_1 y_1 + \alpha_0 =0$
and the affine Hecke algebra is usually called the $B$-type Hecke algebra.
We rewrite the characteristic identity in a form (to simplify notation we set $y:= y_1$)
$$
\frac{1}{y - \xi \, x^{-1}} =
- \frac{(y+\alpha_1 + \xi \, x^{-1})}{(\xi^2 x^{-2} + \alpha_1 \xi x^{-1} + \alpha_0)} \; .
$$
A substitution of this equation in (\ref{soluH}) gives the simplest
polynomial solution \cite{DoiLevi}:
\be
\lb{marlev}
y(x) = \frac{(x - x^{-1})}{(\xi \, x^{-2} + \alpha_1 x^{-1} + \alpha_0/\xi)}
\left(y + \frac{x \alpha_1 +\xi + \alpha_0/\xi}{x-x^{-1}}  \right) \; .
\ee

\vspace{0.2cm}
\noindent
\underline{Case $m=2$.} In this case the identity (\ref{chart}) is of order 3
and the formula (\ref{kulm2}) gives a solution of the reflection equation \cite{MudK}
$$
\!\!\!\! y(x) \! = \! \frac{\xi(x - x^{-1})}{(\frac{\xi}{x})^3  +
\alpha_2 (\frac{\xi}{x})^2+ \alpha_1 \frac{\xi}{x} + \alpha_0}
\left( \! y^2 + \left( \! \frac{\xi}{x}+\alpha_2 \! \right) y +
\frac{\frac{\xi^2}{x} +\alpha_2 \xi + \alpha_1 x  + \frac{\alpha_0}{\xi}}{x-x^{-1}} \! \right) \!
$$
\be
\lb{m2}
= \frac{\xi(x - x^{-1})}{\sum_{r=0}^{3} \, \alpha_r (\frac{\xi}{x})^r}
\left( \frac{\xi}{x} \,  y +
\frac{\frac{\xi^2+\alpha_1}{x} +\alpha_2 \xi  + \frac{\alpha_0}{\xi}}{x-x^{-1}}
- \frac{\alpha_0}{y} \right) , \;\;\; (\alpha_3 :=1) .
\ee

\vspace{0.2cm}
\noindent
\underline{Case $m=3$.} The characteristic identity is of order 4 and
the solution (\ref{kulm2}) takes the form
$$
\begin{array}{l}
\!\!\!\!\!\!\! y(x) = \frac{\xi(x - x^{-1})}{\sum_{r=0}^{4} \, \alpha_r (\frac{\xi}{x})^r}
\left( \! y^3 + \left( \!  \frac{\xi}{x}+\alpha_3 \! \right) y^2 +
\left( \! \frac{\xi^2}{x^2}+ \alpha_3 \frac{\xi}{x} + \alpha_2 \! \right) y  \right. \\ \\
\left. + \frac{x(\frac{\xi^3}{x^3} +\alpha_3
\frac{\xi^2}{x^2} +\alpha_2 \frac{\xi}{x} + \alpha_1)  + \frac{\alpha_0}{\xi}}{x-x^{-1}} \! \right) ,
\;\;\; (\alpha_4 :=1) \; .
\end{array}
$$

\vspace{0.2cm}
\noindent
{\bf Remark 4.} Let $V$ be an $N$-dimensional vector space.
In an $R$-matrix representation $\rho_R$: $\hat{H}_{M+1} \to {\rm End}(V^{\otimes (M+1)})$ of
the affine Hecke algebra $\hat{H}_{M+1}$,
\be\lb{Rref1}
\rho_R(\sigma_n) = \hat{R}_{n}\; , \;\;\rho_R(y_n) = L_n \; ,
\ee
one has $(n=1, \dots , M)$
\be
\hat{R}_{n} \,  \hat{R}_{n+1} \,\hat{R}_{n} =
\hat{R}_{n+1} \,  \hat{R}_{n} \,\hat{R}_{n+1} \; , \;\;\;
\ee
\be\lb{refA1-1}
\hat{R}_{1} \, L_1 \, \hat{R}_{1} \, L_1 =
L_1 \, \hat{R}_{1} \, L_1 \, \hat{R}_{1} \; ,
\ee
\be
\lb{refAl}
L_{n+1} = \hat{R}_{n} \, L_n \hat{R}_{n},
\ee
\be\lb{refA1bis}
\hat{R}_{n}:=\hat{R}_{n \, n+1} = I^{\otimes (n-1)} \otimes \hat{R} \otimes I^{\otimes (M-n)}
\in {\rm End}(V^{\otimes (M+1)}) \; ,
\ee
where $\hat{R} \in {\rm End}(V \otimes V)$ is the Hecke type
$R$-matrix,
$\hat{R}^2 = \lambda \hat{R} +I^{\otimes 2}$,
 and $I$ is an identity matrix in ${\rm End}(V)$.
The operator $L_1 = \rho_R(y_1)$ is taken in a form
$L_1 = L \otimes I^{\otimes M}$
where $L$ is a $N \times N$ matrix whose entries are generators
of a unital associative algebra ${\cal A}$ which is called
{\it reflection equation} algebra.

\medskip
We recall that for the Hecke algebra one can define inductively a
set of antisymmetrizers \cite{Jimb1}:
\be
\lb{antirr}
A_{1 \to 1}=1  \; , \;\;\; A_{1 \to k+1} =
A_{1 \to k}
\, \widetilde{\sigma}_{k}( a^{2k} ;a) \, A_{1 \to k} \; , \;\;\; (k = 1,2, \dots ) \; ,
\ee
where $\widetilde{\sigma}_k(x;a)$ is the unitary baxterized element (\ref{baxtH2}) and
$a=q$ (for $a=-q^{-1}$ eqs.(\ref{antirr})
define the set of symmetrizers). The Hecke type
$R$-matrix is called the $R$-matrix of height $(m+1)$ if
$\rho_R(A_{1 \to m+2})=0$
and the operator $\rho_R(A_{1 \to m+1})$ has rank 1. We also require that the $R$-matrix
is skew invertible, that is, $\exists\  F \in {\rm End}(V \otimes V)$
such that $Tr_3(F_{13} R_{32})= P_{12}$, where $P_{12} \in {\rm End}(V \otimes V)$
is a permutation matrix. For such $R$-matrices
(e.g., the $U_q(gl(m+1))$ Drinfeld-Jimbo $R$-matrix in the vector representation is the Hecke type $R$-matrix
of height $(m+1)$),
it was shown in \cite{GPS} (see also \cite{IOPS} and references therein)
that the quantum matrix $L$ of the reflection equation algebra ${\cal A}$
satisfies the characteristic
identity of order $(m+1)$ whose coefficients are central in ${\cal A}$.
It means that in the $R$-matrix representation (\ref{Rref1}) (with the skew invertible $R$-matrix
of finite height)
the affine Hecke algebra is effectively the cyclotomic one (over the center) and hence the
general solutions (\ref{soluH}) of the reflection equation (\ref{reflH})
are reduced to the polynomial solutions (\ref{kulm2}).

\vskip .2cm
We note also that a Temperley-Lieb algebra
is a quotient of the Hecke algebra by a relation $A_{1 \to 3}=0$. Hence the
$R$-matrix representation of the Temperley-Lieb algebra
is associated to the Hecke type $R$-matrix of height $2$. In this case
(e.g. in the formulation of integrable XXZ spin chain models with nontrivial
boundary conditions) it is enough to
use the simplest polynomial solution (\ref{marlev}).

\vspace{0.2cm}
\noindent
{\bf Remark 5.} Consider the $R$-matrix representation $\rho_R$ of the
affine Hecke algebra $\hat{H}_{M+1}$ (\ref{Rref1}),
where $\hat{R} \in {\rm End}(V^{\otimes 2})$ is the standard Drinfeld-Jimbo
$R$-matrix for $U_q(gl(N))$:
\be\label{drjr}
\hat{R}=\sum_i  \, q \, e_{ii} \otimes e_{ii} +
 \sum_{i \neq j}  \, e_{ij} \otimes e_{ji} +
\lambda \,  \sum_{j > i} \, e_{ii} \otimes e_{jj} \; ,
\ee
$$
\hat{R}^2 =\lambda \hat{R} + I^{\otimes 2} \; , \;\;\;
\lim_{q \to 1} \left( \hat{R}_{12} \right) = P_{12} \; .
$$
Here $e_{ij}$ are matrix units and $P_{12}$ is the permutation matrix.
The $R$-matrix (\ref{drjr}) is skew invertible Hecke type $R$-matrix of height $N$.

\medskip
Let $L^{+}$ (respectively, $L^{-}) \in {\rm Mat}(N)$
be upper (respectively, lower) triangular operator-valued invertible matrices
which satisfy:
\be
\lb{frt}
\begin{array}{c}
\hat{R}_{12} \,  L^+_2 \,  L^+_1 =
L^+_2 \,  L^+_1  \, \hat{R}_{12}  \; , \\[0.2cm]
\hat{R}_{12} \,  L^-_2 \,  L^-_1 =
L^-_2 \,  L^-_1  \, \hat{R}_{12}  \; , \\[0.2cm]
\hat{R}_{12} \,  L^{+}_2 \, L^{-}_1 = L^{-}_2 \,  L^{+}_1 \,  \hat{R}_{12} \; .
\end{array}
\ee
According to the approach of \cite{10},
$L^{\pm}$ are matrices of Cartan type generators of $U_q(gl(N))$.
It is known that the formula
\be
\lb{rlml}
\rho_R(y_1) = L_1 := (1/L^{-}) \,  L^{+} \otimes I^{\otimes M}
\ee
defines a solution of the reflection equation (\ref{refA1-1}).

\smallskip
For the operator (\ref{rlml}), the solution (\ref{soluH})
for $n=1$ takes the form
\be
\lb{rhoy}
\rho_R(y_1(x)) =
\frac{1}{(L^+_1 - \xi x^{-1} L^-_1)}(L^+_1 - \xi x L^-_1) =: K_1(x)\; .
\ee
The operator $K(x)$ in the formula (\ref{rhoy})
is represented in the factorized form $L^{-1}(\xi/x) L(\xi x)$, where the
operator-valued matrix
\be
\lb{eval}
L(x) = (L^+ - x L^-) \; ,
\ee
defines the evaluation representation (see, e.g., \cite{ChPr})
of the affine algebra $U_q(\hat{gl}(N))$ and
satisfies the intertwining relations (see, e.g., \cite{Isa})
\be
\lb{intertw}
\hat{R}_{12}(x) \, L_2(x y)  \, L_1(y) =
L_2(y)  \, L_1(x y) \, \hat{R}_{12}(x) \; .
\ee
Using (\ref{intertw}) one can immediately check that (\ref{rhoy}) solves the reflection equation
(\ref{reflH}) for $n=1$ written in the $R$-matrix form
\be
\lb{reflHr}
\hat{R}_{12} (x/z) \, K_1(x) \,
\hat{R}_{12}  (x \, z) \, K_1(z) =
K_1(z)\, \hat{R}_{12} (x \, z) \, K_1(x) \,
\hat{R}_{12}  (x/z) \; ,
\ee
where
$$
\hat{R}_{12} (x):= \hat{R}_{12} - x \hat{R}_{12}^{-1} = \rho_{R}(\sigma_1(x)) \; ,
$$
is the $R$-matrix
image of the baxterized element (\ref{baxtH}).

\section{Solutions of reflection equations \\ for the BMW algebra}\lb{bmwsection}
\setcounter{equation}0

In this section, we search for solutions of the reflection equations (\ref{reflH}) for the
Birman-Murakami-Wenzl (BMW) algebra. The BMW algebra $BMW_{M+1}$ is generated
by elements $\sigma_n$ (\ref{braidg})
and elements $\kappa_n$ $(n=1,\dots,M)$ which
satisfy following relations \cite{W2}:
\be
\lb{bmw1}
\kappa_n \, \sigma_n =  \sigma_n \, \kappa_n = \nu \, \kappa_n \; ,
\ee
\be
\lb{bmw3}
\sigma_n - \sigma_n^{-1} = \lambda \, ( 1 - \kappa_n) \; ,
\ee
\be
\lb{bmw2}
\kappa_n \, \sigma_{n-1}^{\pm 1} \, \kappa_n = \nu^{\mp 1} \, \kappa_n \; ,
\ee
where $\lambda = q -q^{-1}$ and
$\nu \in {\bf C}\backslash \{0, \pm q^{\pm 1} \}$
is an additional parameter of the algebra. The BMW algebra is a finite dimensional
quotient of the group algebra of the braid group ${\cal B}_{M+1}$.

\medskip
For the BMW algebra the baxterized elements (which solve the Yang-Baxter equation (\ref{ybeH}))
have the form  \cite{Jon2}, \cite{Isa}:
\be\begin{array}{c}
\sigma_n(x)= {\displaystyle\frac{1}{\nu+a x^{-1}}}
\left( a \, (x^{-1}-1) \, \sigma_n + \nu \, (1-x) \, \sigma_n^{-1} +
\lambda \, (a + \nu) \right)
\\[1em]
\lb{bmwbax}
 = ( \sigma_n - x \, \sigma^{-1}_n) +
\lambda \, {\displaystyle\frac{(\nu +a)}{( \nu + a x^{-1})}} \, \kappa_n =
\left( a \, x^{-1} - a^{-1} \right)
{\displaystyle\frac{\sigma_n + a \, x}{\sigma_n  + a \,  x^{-1}}} \; ,
\end{array}\ee
where $x$ is a spectral parameter and $a$ is a solution of the equation
$\lambda = a  - a^{-1}$, i.e. $a=\pm q^{\pm 1}$.
The last expression in (\ref{bmwbax}) coincides
with the form (\ref{baxtH1}) of the baxterized element for the Hecke algebra.
However, in the case of the BMW algebra, two different possible values of $a=\pm q^{\pm 1}$
lead to two different baxterized solutions.

\medskip
We consider the affine algebra $\overline{BMW}_{M+1}$ with
generators $\sigma_n$, $\kappa_n$, $(n=1,\dots,M)$, $y_n$
$(n=1,\dots,M+1)$, $c$ and $Q^{(k)}$ ($k$ is a positive integer) subject
to the relations (\ref{braidg}), (\ref{afheck}), (\ref{bmw1}) -- (\ref{bmw2}) together with
\be
\lb{bmwa5}
\kappa_1 y_1 \sigma_1 y_1 \sigma_1 = c \, \kappa_1 = y_1 \sigma_1 y_1 \sigma_1 \, \kappa_1
\; ,
\ee
\be
\lb{bmwa6}
\kappa_1 y^k_1 \kappa_1 = Q^{(k)} \, \kappa_1  \; .
\ee
The generators $Q^{(k)}$ and $c$ are assumed to be central. We additionally suppose that
the elements $y_1$ and $c$ are invertible and moreover the element
$c$ has a square root in the algebra $\overline{BMW}_{M+1}$.
It is natural to introduce elements $Q^{(k)}$ with $k\leq 0$, which satisfy the relation
(\ref{bmwa6}) as well; we have $Q^{(0)}= (\nu^{-1} +\lambda - \nu)/\lambda$; the elements
$Q^{(k)}$ with $k< 0$ are expressed in terms of $Q^{(k)}$ with $k>0$ by
\be
\lb{nazar}
Q^{(-n)} = \nu^{2} \, c^{-n} \,\, Q^{(n)} +  \lambda  \, \nu \, \sum_{j=1}^{n-1} \, c^{-j}
\left( Q^{(2j-n)} - Q^{(j)}\, Q^{(j-n)} \right)  \; .
\ee
A subalgebra generated by the elements
$\sigma_n$ and $\kappa_n$ is the usual BMW algebra $BMW_{M+1}$.
The abelian subalgebras which are generated by sets $\{ c, Q^{(k)},y_1 \}$
and $\{ c, Q^{(k)} \}$ can be denoted $\overline{BMW}_{1}$ and $\overline{BMW}_{0}$,
respectively. Note that our
definition of the affine BMW algebra is slightly different from the definition
of the affine BMW algebra accepted in \cite{Ram} and \cite{RO},
where the central elements $Q^{(k)}$ and $c$
are taken to be constants (see, however, the definition
of a "degenerate affine Wenzl algebra" in \cite{Nazar}).

\medskip
The affine Hecke algebra $\hat{H}_{M+1}$ is isomorphic to a subalgebra, generated
by the elements $\sigma_n$ and $y_n$, in the quotient of the affine BMW algebra
$\overline{BMW}_{M+1}$ by an ideal generated by $\{ \kappa_n \}$.
Having this in mind, it is natural to look for a solution
of the reflection equation (\ref{reflH})
(for the BMW algebra case) in the form (\ref{soluH}).
However, contrary to the case of the affine Hecke algebra,
the algebra $\overline{BMW}_{M+1}$ does not possess the automorphism $y_n \to \zeta y_n$
(see Remark 2 of the previous Section) because of the relations (\ref{bmwa5}).
So we expect that the parameter $\xi$ cannot be a free
parameter of the solution (\ref{soluH}) and should be fixed in the BMW algebra case.
Indeed, we have the following result.

\vspace{0.2cm}
\noindent
{\bf Proposition 2.}
{\it For the affine BMW algebra a
local unitary solution of the reflection equation (\ref{reflH})
(with $\sigma_n(x)$ defined in (\ref{bmwbax})) is
\be
\lb{solBMW}
y_n(x) = f(x) \, \frac{y_n - \xi \, x }{y_{n} - \xi \, x^{-1}} \; ,
\ee
where $\xi$ is a central element in $\overline{BMW}_{M+1}$
fixed by $\xi^2 = -a \, c /\nu$ and $f(x)$ is a
scalar function such that $f(x)f(x^{-1})=1$.
}

\vspace{0.2cm}
\noindent
{\bf Proof.} Formula (\ref{solBMW})
can be checked by quite lengthy brute force calculations. \hfill $\bullet$

\medskip
Without loss of generality we will consider the equation (\ref{reflH})
and its solutions (\ref{solBMW}) for $n=1$.
In this case, it is sufficient to consider an affine subalgebra $\overline{BMW}_2$ with
generators $\sigma_1$, $\kappa_1$, $y_1$, $c$ and $Q^{(k)}$ $(k>0)$.
In addition to $\{ c,Q^{(k)} \}$,
the center of the algebra $\overline{BMW}_2$ contains elements
$(y_1 y_2)$ and $f(y_1,y_2)(c - y_1 \, y_2)$,
where $f(y_1,y_2)$ is any symmetric
function of $y_1$ and $y_2 := \sigma_1 y_1 \sigma_1$.

\medskip
One can use (\ref{solBMW}) to construct solutions for a cyclotomic
affine BMW algebra when the generator $y_1$ satisfies a polynomial characteristic identity
(\ref{chart}). It can be done by the method which was explained in
Remark 3 of the previous Section. Solutions are given by the
same formula (\ref{kulm2}), where the parameter $\xi$ should be fixed as in
Proposition 2.
It is worth noting that in the BMW case the parameters $\alpha_k$ in (\ref{chart})
are not all independent; they are related by conditions
\be
\lb{chart1}
Q^{(m+1-r)} + \sum_{k=0}^{m} \alpha_k Q^{(k-r)} =0 \;\;\; (r=0,1,\dots,m+1) \; ,
\ee
which follow from (\ref{chart}).

\vspace{0.3cm}
The solution (\ref{solBMW}) is not general.
Below we present two exceptional solutions of (\ref{reflH}) (for $n=1$)
for cyclotomic affine BMW algebras
of orders 2 and 4. In general these solutions cannot be produced from (\ref{solBMW}).
To simplify formulas, we omit all scalar
factors which are needed for the unitarity of $y_1(x)$ and use
a concise notation $y:=y_1$, $y(x):=y_1(x)$.

\vspace{0.2cm}
\noindent
{\bf (1)} For the characteristic identity of degree 2,
\be
y^2 + \alpha_1 \, y + \alpha_0 =0 \; ,
\ee
the compatibility conditions (\ref{chart1}) imply relations between the elements $Q^{(k)}$,
$\alpha_0$ and $\alpha_1$.
First,
\be
\lb{nnww}
\alpha_1=-\frac{\lambda (c+\nu^2\alpha_0 )}{c(\nu^{-1}-\nu +\lambda )}Q^{(1)}\ .
\ee
Then we have two possibilities.

\smallskip
\noindent
{\bf (1a)} The element $Q^{(1)}$ is free while the values of the
parameters $\alpha_0$ and $\alpha_1$ become fixed:
\be\lb{fix1}
\alpha_0 = -\frac{c}{a \, \nu} \; , \;\; \alpha_1 = - \frac{Q^{(1)} \, \nu \, \lambda}{a \nu +1} \; .\ee
Due to the characteristic identity, the elements $Q^{(k)}$ with $k>1$ can be expressed in terms of
$Q^{(1)}$.

\smallskip
We have in this case the same solution as in (\ref{marlev}):
\be\lb{marlev2}
\begin{array}{c} y(x) = y + \frac{\alpha_1 \, x + A}{x-x^{-1}} \; , \;\;\;\end{array}\ee
where $A$ is an arbitrary constant.

Here the value of the parameter $a$ ($a=q$ or $a=-q^{-1}$) should be the same as in
the definition of the baxterized element (\ref{bmwbax}) which enters
the reflection equation (\ref{reflH}). The special case of (\ref{marlev2}) with
$A= -\frac{c}{\xi \nu} (a +1/a)$, $(\xi^2= -a \, c /\nu)$, can be produced
from the rational solution (\ref{solBMW}). Another special example of (\ref{marlev2})
was presented in \cite{RO}.

\smallskip
\noindent
{\bf (1b)} The values of the elements $Q^{(k)}$ are fixed,
\be Q^{(k)}=\frac{c^{k/2}}{\nu^k}\ Q^{(0)}\ .\ee
In this case the parameter $\alpha_0$ is free, the characteristic polynomial for $y$
factorizes, $(y-c^{1/2}/ \nu )(y-\alpha_0\nu /c^{1/2})=0$, and the solution (\ref{solBMW}) produces the
solution (\ref{marlev2}) with $A=\xi +\alpha_0/ \xi$.

\vspace{0.2cm}
\noindent
{\bf (2)} For the characteristic identity of degree 4,
\be
\lb{sol4}
 y^4 + \alpha_3 \,  y^3  + \alpha_2 \,  y^2 + \alpha_1 \,  y + \alpha_0 = 0 \; ,
\ee
the formula
\be
\lb{sol5}
y(x) = \bar{\alpha}_0 \, y + \frac{\alpha_3 \, \bar{\alpha}_0 \, x + \alpha_1}{(x-x^{-1})} -
x \, \alpha_0 \, \frac{1}{y} \; ,
\ee
where $\bar{\alpha}_0^2 =\alpha_0$, defines a solution of the reflection equation if
$$
\alpha_0 = - \frac{c^2}{a \, \nu} \; , \;\;
\alpha_1 =  -\frac{c}{a} \left( \frac{\alpha_3}{\nu}  + \lambda Q^{(1)} \right)\; ,
$$
\be
\lb{sol6}
\alpha_2 = - \frac{\nu \lambda}{a \nu+1} \left(Q^{(1)} \, \alpha_3 + Q^{(2)} \right)
+ \frac{\lambda c}{a} \; ,
\ee
$$
\alpha_3 = - \, \frac{Q^{(3)} -\frac{\nu \lambda  Q^{(1)}  Q^{(2)}}{a\nu +1} -\frac{c}{\nu a} \, Q^{(1)}
}{ \left( Q^{(2)} - \frac{\nu \lambda (Q^{(1)})^2}{(a\nu+1)}
- \frac{c}{\nu a} \, Q^{(0)} \right)} \; ,
$$
(here $Q^{(0)} = (1/\nu + \lambda -\nu)/\lambda$). The
choice of the parameters $\alpha_i$ in (\ref{sol6}) is consistent with the constraints (\ref{chart1}).

\smallskip
The solution (\ref{sol5})
has the form of the special {\it small} solution of the reflection
equation in the Hecke algebra case (see \cite{MudK}).

\vskip .2cm
\noindent
{\bf Remark 6.}
The $R$-matrix representation of the affine BMW algebra is defined by eqs.
(\ref{Rref1})--(\ref{refA1bis}) and the corresponding matrix versions of relations
(\ref{bmw1})--(\ref{bmwa6}). As for the Hecke algebra case, one can define a set
of antisymmetrizers for the BMW algebra \cite{OgPya}, \cite{Isa}, \cite{TW}
\be
\lb{antirBMW}
A_{1 \to 1}=1  \; , \;\;\; A_{1 \to n+1} =
A_{1 \to n}
\; \widetilde{\sigma}_{n}( a^{2n};a ) \; A_{1 \to n} \; \; \;\;\; (n = 1,2, \dots ) \; ,
\ee
where the unitary element $\widetilde{\sigma}_{n}(x;a)$ is given by (\ref{baxtH2}) with the BMW
baxterized element (\ref{bmwbax}) and $a=q$ (for $a=-q^{-1}$ eqs.(\ref{antirBMW})
define the set of symmetrizers).
If the BMW type $R$-matrix (additional information about
the BMW type $R$-matrices can be found in \cite{10}, \cite{Isa}, \cite{IOP5}) is skew invertible,
$\rho_R(A_{1 \to m+1})=0$ and $\rho_R(A_{1 \to m})\neq 0$, then
the corresponding quantum matrix $L$ of the reflection equation algebra ${\cal A}$ (\ref{refA1-1})
satisfies the characteristic
identity of order $m$ whose coefficients are central in ${\cal A}$ \cite{OgPya}
(see also \cite{Mud}).
In this case, for
the solutions of the reflection equation (\ref{reflH}), one
can use (\ref{marlev2}), (\ref{sol5}) and the polynomial solutions  of the type
(\ref{kulm2}), where $\xi^2=-ac/\nu$ and $y_1$ is the generator of the
cyclotomic affine BMW algebra.

\section{Relations to integrable spin chain models}
\setcounter{equation}0

Consider the reflection equation (\ref{reflH}) in the generic case when
$\sigma_n(x)$ is a baxterized element of the group algebra of ${\cal B}_{M+1}$
(or its quotients). In this case, we have the following statement:

\vspace{0.2cm}
\noindent
{\bf Proposition 3.} {\it Let $y_1(x)$ be a local  (i.e., $[y_1(x),\sigma_m]=0$ $\forall m>1$)
solution of (\ref{reflH}) for $n=1$. Then
\be
\lb{xxz5}
\bar{y}_n(x) = \sigma_{n-1}(x) \cdots \sigma_1(x) \,
y_1(x) \, \sigma_{1}(x) \cdots \sigma_{n-1}(x) \; ,
\ee
is a (non-local) solution of the reflection equation (\ref{reflH}) for $n>1$.}

\vspace{0.2cm}
\noindent
{\bf Proof.} We prove (\ref{xxz5}) by induction.
The element in the right hand side of eq.(\ref{xxz5}) commutes with $\sigma_m$ for $m>n$.
Let $\bar{y}_{k-1}(x)$ be a solution of
(\ref{reflH}) for $n=k-1$ and  $[\bar{y}_{k-1}(x),\sigma_m]=0$ $(m>k-1)$.
Then, using (\ref{ybeH}) and (\ref{reflH})
with $n=k-1$, we obtain that
\be
\lb{xxz5a}
\bar{y}_k(x) = \sigma_{k-1}(x) \, \bar{y}_{k-1}(x) \, \sigma_{k-1}(x) \; ,
\ee
solves (\ref{reflH}) for $n=k$.
\hfill $\bullet$

\medskip
Consider a direct product:
$\hat{\hat{H}}_{M+1} = \hat{H}_{M+1} \otimes \hat{\hat{H}}_{0}$ of the affine Hecke algebra
$\hat{H}_{M+1}$ and an abelian algebra $\hat{\hat{H}}_{0}$ generated by
commutative elements $Q_D^{(k)}$ $(k \in {\bf Z})$. The algebras $\hat{\hat{H}}_{M+1}$
admit a chain of inclusions
$$
\hat{\hat{H}}_{0} \subset \hat{\hat{H}}_{1} \subset \hat{\hat{H}}_{2} \subset \dots
\hat{\hat{H}}_{M} \subset \hat{\hat{H}}_{M+1}
$$
defined on the generators as
$$
\hat{\hat{H}}_{n} \ni ( Q_D^{(k)},y_1, \sigma_i) \longrightarrow
( Q_D^{(k)},y_1, \sigma_i)  \in  \hat{\hat{H}}_{n+1} \;\;\; (i=1, \dots,n-1) \; ;
$$
the algebra $\hat{\hat{H}}_{1}$ is generated by elements $(Q_D^{(k)},y_1)$.
We equip the algebra $\hat{\hat{H}}_{M+1}$
with linear mappings $Tr_{D(n+1)}$: $\hat{\hat{H}}_{n+1} \to \hat{\hat{H}}_{n}$
(from the algebra $\hat{\hat{H}}_{n+1}$ to its subalgebra $\hat{\hat{H}}_{n}$)
such that $\forall X,Y \in \hat{\hat{H}}_{n}$ and $\forall Z \in \hat{\hat{H}}_{n+1}$ we have
\be
\label{map}
\begin{array}{c}
\!\!\!\! Tr_{D(n+1)} ( X \, Z \, Y ) = X \, Tr_{D(n+1)}(Z) \, Y   , \\[0.15cm]
Tr_{D(n+1)} ( \sigma_n^{\pm 1} X \sigma_n^{\mp 1}) =
Tr_{D(n)} (X) \, , \\[0.15cm]
Tr_{D(n+1)} ( 1 ) = D_n^{(0)} \; , \;\;\;
Tr_{D(n+1)} (\sigma_n^{\pm 1}) = D_n^{(\pm 1)} \; , \\[0.15cm]
Tr_{D(1)} (y_1^k)= Q_D^{(k)} \; , \;\; Tr_{D(n)}  Tr_{D(n+1)} ( \sigma_n Z ) =
Tr_{D(n)}  Tr_{D(n+1)} ( Z \sigma_n) \; ,
\end{array}
\ee
where $D^{(m)}_n \in {\bf C}\setminus \{ 0\}$ are constants subject to
the relation
$D^{(+1)}_n = D^{(-1)}_n + \lambda D^{(0)}_n$ which follows
from the Hecke condition (\ref{ahecke}).
In the $R$-matrix representation of $\hat{\hat{H}}_{M+1}$ (\ref{Rref1}) -- (\ref{refA1bis})
the maps $Tr_{D(n+1)}$ (when $D^{(m)}_n$ do not depend on $n$) are nothing but a quantum trace.

\vspace{0.2cm}
\noindent
{\bf Proposition 4.} {\it The
following identity (cross-unitarity)
holds for the baxterized element $\sigma_n(x)$ (\ref{baxtH}):
\be
\lb{map1}
\!\!\!\! Tr_{D(n+1)} \left( \sigma_n(x) \, Y_n \, \sigma_n(b_n/x) \right)  =
\eta(x) \, \eta(b_n/x) \,
Tr_{D(n)} (Y_n) \;  \;\; (\forall \, Y_n \in \hat{\hat{H}}_{n}) \;  ,
\ee
where
$\eta(x) := (1-x)$, $b_n := D^{(+1)}_n/D^{(-1)}_n$.
}

\vspace{0.2cm}
\noindent
{\bf Proof.} Using the definition (\ref{map})
of the map $Tr_{D(n+1)}$ and the explicit form of the baxterized
element (\ref{baxtH}) we obtain
for the left-hand side of (\ref{map1})
$$
 Tr_{D(n+1)} \left( \left((1-x)\sigma_n + \lambda x \right) \; Y_n \;
((1-b_n\, x^{-1})\sigma_n^{-1} + \lambda )\right)
$$
$$
= (1-x)(1-b_n\, x^{-1}) Tr_{D(n+1)}  \left(\sigma_n Y_n \sigma_n^{-1} \right)  +
 \lambda \left( D^{(+1)}_n - b_n\, D^{(-1)}_n \right)  Y_n \; ,
$$
which is equivalent to (\ref{map1}). \hfill $\bullet$

\vspace{0.2cm}
\noindent
{\bf Proposition 5.} {\it Let an element $\bar{y}_n(x)\in \hat{\hat{H}}_{n}$ be a solution
of (\ref{reflH}) (e.g., the solution defined in (\ref{xxz5})). Then
elements $\tau(x) \in \hat{\hat{H}}_{n-1}$
\be
\lb{trmat1}
\begin{array}{c}
\tau (x) = Tr_{D(n)} \left( \bar{y}_n(x) \right) \; ,
\end{array}
\ee
form a commutative family, $[\tau(x), \, \tau(z)]=0$ $(\forall x,z)$.
}

\vspace{0.2cm}
\noindent
{\bf Proof.} Using properties (\ref{map}), the identity (\ref{map1}) and
the reflection equation (\ref{reflH}) we deduce
$$
\tau(x) \, \tau(z) = Tr_{D(n)} \left( \bar{y}_n(x) \, \tau(z) \right)
$$
$$
= \frac{1}{\eta_n(xz)}
Tr_{D(n,n+1)} \left( \bar{y}_n(x) \sigma_n(x z) \,
\bar{y}_n(z)\, \sigma_n(b_n /(x z)) \right)
$$
$$
 = \frac{1}{\eta_n(xz)} Tr_{D(n,n+1)} \left( \sigma_n^{-1}(x/z)  \bar{y}_n(z) \sigma_n(x z) \,
\bar{y}_n(x)\, \sigma_n(b_n /(x z))  \sigma_n(x/z) \right)
$$
$$
 = \frac{1}{\eta_n(xz)} Tr_{D(n,n+1)} \left( \bar{y}_n(z) \sigma_n(x z) \,
\bar{y}_n(x)\, \sigma_n(b_n /(x z))\right) = \tau(z) \, \tau(x) \; ,
$$
where $Tr_{D(n,n+1)}:= Tr_{D(n)} Tr_{D(n+1)}$ and $\eta_n(x) = \eta(x)\eta(b_n/x)$.
This proves the commutativity of the family $\tau(x)$. \hfill $\bullet$

\medskip
In the $R$-matrix representation (\ref{Rref1}) -- (\ref{refA1bis}), the element $\tau(x)$
(with $\bar{y}_n(x)$ taken in the form (\ref{xxz5}))
is nothing but an example of a Sklyanin monodromy matrix \cite{Skl} in the case when
the algebra ${\cal T}_+$ from \cite{Skl} is realized trivially, ${\cal T}_+=1$. Thus, the
element $\tau(x)$ can be used for a
formulation of integrable chain models with nontrivial boundary conditions
on one end of the chain. We consider two possibilities for the solution $y_1(x)$
in (\ref{xxz5}): (1) $y_1^{(1)}(x)=1$ and (2) $y_1^{(2)}(x) = (y_1 - \xi x)(y_1 - \xi/x)^{-1}$
(see (\ref{soluH})). In these cases the corresponding monodromy elements and
local Hamiltonians belong to the affine Hecke algebra $\hat{H}_{n-1}$:
\be
\lb{tau1}
\tau^{(1)}(x) = Tr_{D(n)} \left( \sigma_{n-1}(x) \cdots \sigma_2(x) \, \sigma_1^2(x) \,
 \sigma_{2}(x) \cdots \sigma_{n-1}(x) \right)  \; ,
\ee
\be
\lb{tau2}
\tau^{(2)}(x) = Tr_{D(n)} \left( \sigma_{n-1}(x) \cdots \sigma_1(x) \left(
\frac{y_1 - \xi x}{y_1 - \xi /x} \right) \sigma_{1}(x) \cdots \sigma_{n-1}(x) \right)  \; ,
\ee
\be
\lb{ham2}
{\cal H}^{(1)} = \sum_{m=1}^{n-2} \sigma_m + const \; , \;\;\;
{\cal H}^{(2)} = \sum_{m=1}^{n-2} \sigma_m + \frac{\lambda \xi }{y_1 - \xi}  + const  \; .
\ee
The Hamiltonians ${\cal H}^{(1)}$ (resp., ${\cal H}^{(2)}$) are obtained by differentiating
$\tau^{(1)}(x)$ (resp., $\tau^{(2)}(x)$)
with respect to $x$ at the point $x=1$. The Hamiltonian ${\cal H}^{(1)} $, obtained
for the choice $y_1(x)=1$ (which corresponds to free boundary conditions
at both ends of the chain), is known.
The polynomial analogues of the Hamiltonian ${\cal H}^{(2)}$ in the cyclotomic case
(which corresponds to a nontrivial boundary condition on
one end of the chain) can be deduced in a straightforward manner, as in Remark 3 in
Section \ref{heckesection}.

\medskip
To formulate integrable chain systems with nontrivial boundary conditions on both
ends of the chain, we need to introduce a "conjugated" reflection equation
\be
\lb{crefl'}
\sigma_n(x /z) \, \widetilde{y}_n(z) \, \sigma_n \left(b_n /(x z)\right) \,  \widetilde{y}_n(x) =
\widetilde{y}_n(x) \, \sigma_n(b_n /(x z)) \, \widetilde{y}_n(z) \, \sigma_n(x/z) \; .
\ee
We note that solutions of (\ref{crefl'}) are related to the solutions
of the reflection equation (\ref{reflH}) by means of identities
\be
\lb{isom}
\widetilde{y}_n(x) = y_n \left( \frac{b_n^{1/2}}{x} \right) \;\;\; {\rm or} \;\;\;
\widetilde{y}_n(x) = y_n^{-1} \left( \frac{x}{b_n^{1/2}} \right) \; .
\ee

Now we consider an $R$-matrix representation $\rho_R$ of the algebra $\hat{\hat{H}}_{M+1}$. The
homomorphism $\rho_R$
is described in (\ref{Rref1}) -- (\ref{refA1bis}) and we also have
$\rho_R(Q_D^{(k)}) = Tr_{D(1)}(L_1^k)$, where $Tr_{D}$ is a standard quantum trace
defined for any skew-invertible $R$-matrix \cite{Isa}, \cite{Ogi}. More precisely,
let $F$ be the skew inverse of $\hat{R}$,
$$
Tr_2(F_{12} \hat{R}_{23}) = P_{13} = Tr_2(\hat{R}_{12} F_{23})\ .
$$
Define the operator ${\cal{D}}$ by
$$
{\cal{D}}_1 := Tr_2(F_{12}) \; .
$$
Then
$$
Tr_{{\cal{D}}(n)}(E) := Tr_n({\cal{D}}_n \, E) \; ,  \;\;\;
\forall \, E \in {\rm End}(V^{\otimes (M+1)}) \; .
$$
Here $Tr_n=Tr_{V_n}$ is a usual trace in the copy number $n$ of the space $V$
in the product $V^{\otimes (M+1)}$.
For the Hecke type $R$-matrices,
this definition of the quantum trace implies that the parameters $D^{(\pm 1)}_n$ from (\ref{map}) and
$b_n=D^{(+1)}_n/D^{(-1)}_n$ do not depend on $n$ and are equal to
$$
D^{(+1)}_n=1 \; , \;\;\; D^{(-1)}_n=1-\lambda Tr({\cal{D}}) \; ,
\;\;\; b_n=b := (1-\lambda Tr({\cal{D}}))^{-1} \; .
$$
In a particular example of the standard $R$-matrix of the $GL_q(n|m)$-type we have $b= q^{2(n-m)}$.

\vskip .2cm
The $R$-matrix version of the reflection equation (\ref{reflH}) is
\be
\lb{refl}
\hat{R}_n(x /z) \, K_n(x) \, \hat{R}_n (x z) \,  K_n(z) =
K_n(z) \, \hat{R}_n (x z) \, K_n(x) \, \hat{R}_n(x/z) \; ,
\ee
where $\hat{R}_n(x):= \rho_R(\sigma_n(x))$ and $K_n(x):= \rho_R(y_n(x))$.

\medskip
Let $\widetilde{K}(x)$ be a $(N \times N)$ matrix whose entries are scalar
functions of the spectral parameter $x$:
\be
\lb{scal}
[\widetilde{K}^{i_1}_{j_1}(x),\widetilde{K}^{i_2}_{j_2}(z)]=0 \;\;\;\; (\forall i_1,i_2,j_1,j_2=1, \dots,N)\; .
\ee
Assume that the matrix $\widetilde{K}(x)$ is a
solution of the conjugated reflection equation (\ref{crefl'}) written in
the $R$-matrix representation
\be
\lb{crefl}
\!\!\!\!\!
\hat{R}_n(x /z) \widetilde{K}_n(z) \hat{R}_n \left(b/(x z)\right)  \widetilde{K}_n(x) =
\widetilde{K}_n(x) \hat{R}_n(b/(x z)) \widetilde{K}_n(z) \hat{R}_n(x/z) \; ,
\ee
where $\widetilde{K}_n(z)$ is taken in the local form
\be
\lb{local}
\widetilde{K}_n(z) = I^{\otimes (n-1)} \otimes \widetilde{K}(z) \otimes I^{\otimes (M+1-n)} \; ,
\ee
(we call $\widetilde{K}_n(z)$ local since it acts nontrivially only in the factor $V_n$
in $V^{\otimes(M+1)}$).
We have (cf. Theorem 1 in \cite{Skl}):

\vspace{0.2cm}
\noindent
{\bf Proposition 6.} {\it
Let $K_n(x)$ be a solution of (\ref{refl}) which acts
nontrivially only in the first $n$ factors of $V^{\otimes(M+1)}$.
Assume that the matrix $\widetilde{K}_n(x)$ is a local
scalar solution (\ref{scal}), (\ref{local}) of the conjugated reflection
equation (\ref{crefl}) and all entries of $\widetilde{K}(x)$ commute
with all entries of $K_n(z)$.
Then the elements $t(x)$ defined by
\be
\lb{trmat2}
\begin{array}{c}
t(x) =  Tr_{D(n)} \left( K_n(x) \, \widetilde{K}_n(x) \right)  \;
\end{array}
\ee
form a commutative family, $[t(x), \, t(z)]=0$ $(\forall x,z)$.
}

\vspace{0.2cm}
\noindent
{\bf Proof.} Using properties of the quantum trace
which follow from the definition of the mapping (\ref{map}) and the
$R$-matrix version of the identity (\ref{map1})
we obtain
$$
t(x) \, t(z) = Tr_{D(n)}  \left( {K}_n(x) \,
\left( Tr_{D(n)} ( {K}_n(z) \, \widetilde{K}_n(z) ) \right) \widetilde{K}_n(x) \right)
$$
$$
 = \frac{1}{\eta'(x \, z)} \, Tr_{D(n)} Tr_{D(n+1)} \left( {K}_n(x) \, \hat{R}_n(x \, z) \,
 {K}_n(z) \, \widetilde{K}_n(z) \, \hat{R}_n \left(b/(x z)\right) \widetilde{K}_n(x) \right)\, ,
$$
where $\eta'(x z) = \eta(x z)\eta(b /(xz))$.
Now we apply the reflection equations (\ref{refl}) and (\ref{crefl}) and again use the
properties (\ref{map}) and the identity (\ref{map1})) to rewrite the last expression in the form
$$
 = \frac{1}{\eta'(x \, z)} \, Tr_{D(n)} Tr_{D(n+1)} \left(
{K}_n(z) \, \hat{R}_n(x \, z) \,
{K}_n(x) \, \widetilde{K}_n(x) \, \hat{R}_n \left(b/(x z)\right) \widetilde{K}_n(z) \right)
$$
$$
= Tr_{D(n)}  \left( {K}_n(z) \,
\left( Tr_{D(n)} ( {K}_n(x) \, \widetilde{K}_n(x) ) \right) \widetilde{K}_n(z) \right) =
t(z) \, t(x) \; .
$$
The proof is finished. \hfill $\bullet$

\vspace{0.2cm}

Let $K_n(x)$ in (\ref{trmat2}) be the image of the element $\bar{y}_n(x)$ (\ref{xxz5}) in the
$R$-matrix representation,
$K_n(x)=\rho_R(\bar{y}_n(x))$. With this choice
we obtain the commutative family $t(x)\in {\rm End}(V^{\otimes(n-1)})$:
\be
\lb{trmat}
t(x) = Tr_{D(n)} \left( \hat{R}_{n-1}(x) \cdots \hat{R}_1(x) \,
K_1(x) \, \hat{R}_{1}(x) \cdots \hat{R}_{n-1}(x) \, \widetilde{K}_n(x) \right)  \; ,
\ee
where $K_1(x)$ is any local solution of (\ref{refl}) for $n=1$.
The element $t(x)$ (\ref{trmat}) is an example of
the Sklyanin monodromy matrix \cite{Skl} in the case of
a commutative representation $\widetilde{K}(x)$ of the algebra ${\cal T}_{+}$ \cite{Skl}.
We see that in this case (which is rather general from the point of view of applications)
the proof of the commutativity of $t(x)$ simplifies.

\smallskip
As usual (see \cite{Skl}) one can consider
the element $t(x)$ (\ref{trmat}) as
a generating function for integrals of motion of an integrable open chain
model of length $(n-1)$. If
we take in (\ref{trmat}) the regular solution $K_1(x)$ (i.e. $K_1(1)=1$) of
(\ref{refl}) and take into account
the relation $\hat{R}(1)=\lambda$, we obtain a local Hamiltonian of the chain model
with nontrivial boundary conditions
\be
\lb{ham1}
{\cal H}^{(0)} = \sum_{m=1}^{n-2} \hat{R}_m - \frac{\lambda}{2} \, K_1'(1) +
\frac{Tr_{D(n)}(\hat{R}_{n-1} \, \widetilde{K}_n(1))}{Tr_{D(n)}(\widetilde{K}_n(1))} \; .
\ee
This Hamiltonian is obtained by differentiating $t(x)$ with respect to $x$ at the point $x=1$.
Consider the regular solution $K_1(x)$ of the type (\ref{soluH}),
$$
K_1(x)= \rho_R(y_1(x)) = \frac{L_1 - \xi_1 x}{L_1 -\xi_1 x^{-1}} \; , \;\;\;
(\hat{R}_1 L_1 \hat{R}_1 L_1 = L_1 \hat{R}_1 L_1 \hat{R}_1) \; ,
$$
where $\xi_1$ is a constant.
There are two local solutions $\widetilde{K}_n(x)$ of (\ref{crefl}) which can
be used in (\ref{trmat}) and (\ref{ham1})
\be
\lb{2case}
1) \;\;\; \widetilde{K}_n(x)=1 \; , \;\;\;
2) \;\;\; \widetilde{K}_n(x)=
\frac{\widetilde{L}_n - \xi_2 \, b^{1/2} \, x^{-1}}{\widetilde{L}_n -\xi_2 \, b^{-1/2}\, x} \; .
\ee
Here in the case 2) the matrix
$\widetilde{K}_n(x)$ is chosen according to the equations
(\ref{isom}) and (\ref{soluH}), where $\widetilde{L}_n=\rho_R(y_n)$ is
a scalar and local (see (\ref{scal}) and (\ref{local})) solution of the constant reflection equation
$\hat{R}_n \widetilde{L}_n \hat{R}_n \widetilde{L}_n =
\widetilde{L}_n \hat{R}_n \widetilde{L}_n \hat{R}_n$.

\smallskip
In the first case of (\ref{2case}) the monodromy matrix (\ref{trmat}) and a local Hamiltonian
of the corresponding integrable chain model are given by "$R$-matrix images"
of the elements (\ref{tau1}), (\ref{tau2}), (\ref{ham2}).
In the second case of (\ref{2case}) we obtain an integrable chain model (with
nontrivial boundary conditions for both ends of the chain) with the Hamiltonian
\be
\lb{ham3}
{\cal H}^{(3)} = \sum_{m=1}^{n-2} \hat{R}_m + \frac{\lambda \xi_1}{L_1 - \xi_1} +
\frac{1}{\xi'} \, Tr_{D(n)} \left( \hat{R}_{n-1} \,
\frac{\widetilde{L}_n - \xi_2 \, b^{1/2}}{\widetilde{L}_n - \xi_2 \, b^{-1/2}} \right) \; ,
\ee
where
$\xi' := Tr_{D}\left(\frac{\widetilde{L} - \xi_2 \, b^{1/2}}{
\widetilde{L} - \xi_2 \, b^{-1/2}} \right)$ is a parameter depending on the
choice of the matrix $\widetilde{L}$.
The polynomial analogues of the Hamiltonian (\ref{ham3}) in the cyclotomic case,
when the matrices $L_1$ and
$\widetilde{L}_n$ satisfy the polynomial characteristic
identities of the type (\ref{chart}), can be obtained
straightforwardly, as in Remark 3 in Section \ref{heckesection}.

\medskip
One can generalize the above construction to the case of
the BMW algebra. We do not present here the detailed
description of this construction. The analogue of Proposition 4 reads

\vspace{0.2cm}
\noindent
{\bf Proposition 7.} {\it The
following identity (cross-unitarity)
holds for the BMW baxterized element $\sigma_n(x)$ (\ref{bmwbax})
\be
\lb{map5}
\!\!\!\! Tr_{D(n+1)} \left( \sigma_n(x) \; Y_n \; \sigma_n(z) \right)  =
\eta(x) \, \eta(z) \, Tr_{D(n)} (Y_n) \;\;\; (\forall \, Y_n \in \overline{BMW}_{n}) \;  ,
\ee
where
$$
\eta(x) := (1-x) \, \frac{(a \nu x +1 ) }{(\nu x +a) } \; , \;\;\;
x \, z = \frac{a^2}{\nu^2} \; ,
$$
and the map $Tr_{D(n)}$: $\overline{BMW}_{n} \to \overline{BMW}_{n-1}$
is defined as follows:
$$
 \kappa_{n} Y_n \kappa_n = \frac{1}{\nu} \, Tr_{D(n)}(Y_n) \, \kappa_n \; .
$$
}

\smallskip
The proof of (\ref{map5}) consists in a direct but rather lengthy calculation which we omit.
The conjugated reflection equation for the BMW algebra is given by
(\ref{crefl'}) with $b_n=a^2/\nu^2$. The elements $\sigma_n(x)$ are
the baxterized elements (\ref{bmwbax}). The commutative BMW monodromy elements are given by the same
formulas (\ref{trmat1}) and (\ref{trmat}), where $\bar{y}_n(x)$ is defined in (\ref{xxz5})
with the baxterized elements $\sigma_n(x)$ (\ref{bmwbax});
$R_n(x)$ are the $R$-matrix images of (\ref{bmwbax});
$K(x)$ and $\widetilde{K}(x)$ are the corresponding solutions of
(\ref{refl}) and (\ref{crefl}). The formulas for the commutative monodromy elements (\ref{tau1})
and (\ref{tau2}) (with $\xi^2 = -a \, c/\nu$) are
valid in the BMW case as well.

\medskip
We present the analogues of the
Hamiltonians (\ref{ham2}), (\ref{ham1}) and (\ref{ham3}) in the case of the BMW algebra.
The analogue of ${\cal H}^{(2)}$ (\ref{ham2}) is
\be
\lb{ham5}
{\cal H}^{(4)} = \sum_{m=1}^{n-2} \left( \sigma_m + \frac{\lambda \nu}{\nu +a} \kappa_m \right)
+ \frac{\lambda \xi}{y_1 - \xi}  + const  \; ,
\ee
where $\sigma_m,\kappa_m,y_1$
are generators of the affine BMW algebra, $\xi^2 = -a \, c/\nu$ as in Proposition 2
and the parameter $a$ takes two values $\pm q^{\pm 1}$.
In the case $y_1=1$ we obtain the Hamiltonian (the BMW analogue of ${\cal H}^{(1)}$)
\be
\lb{xxz2}
{\cal H}^{(5)} = \sum_{m=1}^{n-2} \left( \sigma_m + \frac{\lambda \nu}{\nu +a} \kappa_m \right) + const \;
\ee
for the integrable chain models with trivial boundary conditions.
The same Hamiltonian (\ref{xxz2}) defines an integrable system on a periodic chain
if we identify $\sigma_k = \sigma_{n-2+k}$ \cite{Isa}.

\smallskip
The analogue of (\ref{ham1}) is
\be
\lb{ham4}
{\cal H}^{(6)} = \sum_{m=1}^{n-2} \left( \hat{R}_m +  \frac{\lambda \nu}{\nu +a} \hat{\cal K}_m \right)
- \frac{\lambda}{2} \, K_1'(1) +
\frac{Tr_{D(n)}((\hat{R}_{n-1}+ \frac{\lambda \nu}{\nu +a} \hat{\cal K}_{n-1})
\widetilde{K}_n(1))}{Tr_{D(n)}(\widetilde{K}_n(1))} \; ,
\ee
where $\hat{\cal K}_m = \rho_R(\kappa_m)$ and we have two different integrable systems
for $a= \pm q^{\pm 1}$. One can use the solutions (\ref{solBMW}), (\ref{marlev2})
and (\ref{sol5}) in (\ref{ham4}). In particular, one can substitute:
$$
K_1(x)=
\frac{L_1 - \xi \, x}{L_1 -\xi \, x^{-1}} \; , \;\;\;
\widetilde{K}_n(x)=
\frac{\widetilde{L}_n - \xi \, b^{1/2} \, x^{-1}}{\widetilde{L}_n -\xi \, b^{-1/2}\, x}\ ,
$$
where $\xi^2=-a\, c/\nu$, $b=\pm a/\nu$. As a result we deduce the BMW analogue of (\ref{ham3})
$$
{\cal H}_7 = \sum_{m=1}^{n-2} \left( \hat{R}_m +  \frac{\lambda \nu \hat{\cal K}_m}{\nu +a} \right)
+ \frac{\lambda \xi}{L_1-\xi}  + \frac{1}{\xi''}
Tr_{D(n)}\left( \!\! \left(\hat{R}_{n-1}+ \frac{\lambda \nu \hat{\cal K}_{n-1}}{\nu +a} \!\! \right)
\widetilde{K}_n(1) \!\! \right)  ,
$$
where $\xi'' = Tr_{D(n)}(\widetilde{K}_n(1))$ and $\widetilde{K}_n(1) =
(\widetilde{L}_n - \xi \, b^{1/2})(\widetilde{L}_n -\xi \, b^{-1/2})^{-1}$.

\medskip

\noindent
{\bf Remark 7.}
We expect that any representation of the affine Hecke or BMW algebras will give
integrable chain models with the Hamiltonians (\ref{ham2}) or (\ref{ham5}) and (\ref{xxz2}),
respectively. In particular, one can
use in (\ref{ham5}), (\ref{xxz2}) the $SO_q(N)$, $Sp_q(2m)$ and $OSp_q(N|2m)$
$R$-matrix representations of the algebra $BMW_{M+1}$
(see, e.g., \cite{Isa}) or $\overline{BMW}_{M+1}$
to formulate integrable spin chain models of $SO$, $Sp$ or $OSp$ types.
In these cases, the parameter $\nu$ is fixed as follows:
$\nu = q^{1-N}$ for $SO(N)$-type, $\nu = -q^{-1-2m}$
for $Sp(2m)$-type \cite{10}, \cite{OgPya}
and $\nu = q^{1+2m-N}$ for $OSp(N|2m)$-type \cite{Isa}
(there is also a special case $\nu = -q^{-1+2m-2n}$ which corresponds
to $Osp_q(2m|2n)$ \cite{Isa}).
The Yangian limits of the corresponding $R$-matrices
lead to the consideration of $SO$, $Sp$ \cite{28}
or $OSp$ \cite{Arn} invariant spin chain models.
These Yangian models are generalizations of the XXX Heisenberg models of magnets.

\vspace*{-2pt}

\section*{Acknowledgments}

The authors are grateful to P.P. Kulish
for useful discussions and the information about the results presented in \cite{MudK}.
We thank P.N. Pyatov for useful comments.
A.P.I. thanks the Max-Planck-Institut f\"{u}r Mathematik
in Bonn, where part of this work was done, for their kind hospitality
and support. This work was also supported by grants
INTAS 03-51-3350 and RFBR 05-01-01086-a and the ANR project GIMP No.
ANR-05-BLAN-0029-01.

\end{document}